\documentclass[twocolumn,aps,amssymb,footinbib]{revtex4-2}
\usepackage{notoccite}
\usepackage{amssymb}
\usepackage{graphicx}
\usepackage{amsmath}
\usepackage{mathrsfs}
\usepackage{times}
\usepackage{color}
\usepackage{subfigure}
\usepackage{feynmp}
\usepackage{bbold}
\usepackage{bm}
\usepackage{wasysym}
\usepackage{mathtools}
\usepackage{cancel}

\usepackage{xcolor}
\usepackage{float}
\usepackage{microtype}
\usepackage [english]{babel}
\usepackage [autostyle, english = american]{}

\newcommand{\rr}{\mathbf{r}}
\newcommand{\xx}{\mathbf{x}}
\newcommand{\RR}{\mathbf{R}}
\newcommand{\vv}{\mathbf{v}}
\newcommand{\ii}{\mathcal I}

\begin{document}

\begin{abstract}
\noindent 
Active materials take advantage of their internal sources of energy to self-organize in an automated manner. This feature provides a novel opportunity to design micron-scale machines with minimal required control. However, self-organization goes hand in hand with predetermined dynamics that are hardly susceptible to environmental perturbations. Therefore utilizing this feature of active systems requires harnessing and directing the macroscopic dynamics to achieve specific functions; which in turn necessitates understanding the underlying mechanisms of active forces. Here we devise an optical control protocol to engineer the dynamics of active networks composed of microtubules and light-activatable motor proteins. The protocol enables carving activated networks of different shapes, and isolating them from the embedding solution. Studying a large set of shapes, we observe that the active networks contract in a shape-preserving manner that persists over the course of contraction. We formulate a coarse-grained theory and demonstrate that self-similarity of contraction is associated with viscous-like active stresses. These findings help us program the dynamics of the network through manipulating the light intensity in space and time, and maneuver the network into bending in specific directions, as well as temporally alternating directions. Our work improves understanding the active dynamics in contractile networks, and paves a new path towards engineering the dynamics of a large class of active materials.

\end{abstract}

\title{Programming Boundary Deformation Patterns in Active Networks}

\author{Zijie Qu$^{1}$, Jialong Jiang$^{1}$, Heun Jin Lee$^{2}$, Rob Phillips$^{1,2,3}$} 

\author{Shahriar Shadkhoo$^{1}$}
\email{shahriar@caltech.edu}

\author{Matt Thomson$^{1}$}
\email{mthomson@caltech.edu}

\affiliation{$^{1}$Division of Biology and Biological Engineering, California Institute of Technology, Pasadena, CA, USA.}
\affiliation{$^{2}$Department of Applied Physics, California Institute of Technology, Pasadena, CA, USA.}
\affiliation{$^{3}$Department of Physics, California Institute of Technology, Pasadena, CA, USA.}

\maketitle

The rich and exotic dynamical behavior of active systems originates from energy consumption at the level of their constituents, which drives them out of equilibrium and endows them with the capability of self-organizing into micron-scale machines \cite{marchetti2013hydrodynamics,ramaswamy2010mechanics,bechinger2016active,burla2019mechanical}. A central goal is to harness the internally-generated dynamics and programming active stresses to accomplish desired tasks, through modulating the system boundaries and forces at macroscopic scales. Biology has served as the major source of inspiration in designing synthetic active systems \cite{peraza2014origami,pinson2017self,furthauer2020design}. In cells, cross-linked polymer networks mediate the active forces that are generated by motor proteins through hydrolyzing ATP. In vitro experiments with cell extracts and reconstituted networks of Microtubules (MTs) and kinesin motor proteins show self-organization into structures including asters and contractile/extensile networks ~\cite{nedelec1997self,surrey2001physical,sanchez2012spontaneous,foster2015active}. Mechanical properties of active networks have been extensively studied, experimentally \cite{mizuno2007nonequilibrium,thoresen2011reconstitution,kohler2011structure,kohler2012contraction,foster2015active,schuppler2016boundaries,foster2017connecting,suzuki2017spatial} as well as theoretically \cite{lee2001macroscopic,liverpool2003instabilities,kruse2004asters,aranson2005pattern,liverpool2006rheology,juelicher2007active,mackintosh2008nonequilibrium,koenderink2009active,gao2015multiscale_prl,ronceray2016fiber,gladrow2016broken,furthauer2019self}. 

Important questions to be answered include: What modes of dynamics can potentially be probed in a controllable way, and how do we accomplish that? In this paper we address these questions in MT-motor proteins active networks. The interactions of such networks can be categorized into active and passive internal interactions, and network--environment interactions. The latter depend on the specific instrumentation of the experiments, often in an uncontrollable manner. 

\begin{figure*}[ht!]
\centering
\centerline{\includegraphics[width=15cm] {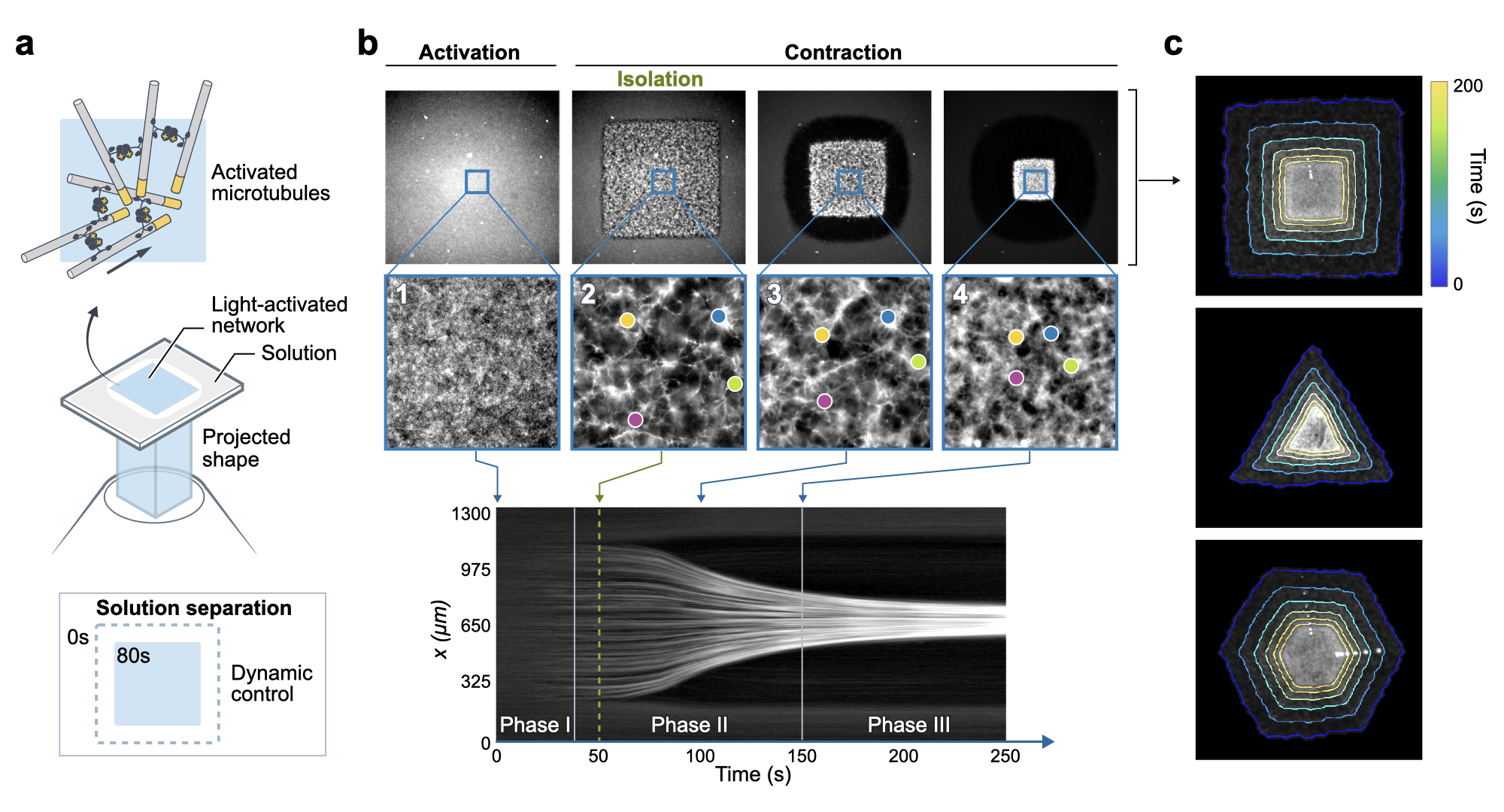}}
\caption{\textbf{Optical-control protocol first activates cross-linking motor proteins to form the MT networks and isolates the network from embedding solution, allowing them to contract self-similarly.} (a) An initial pulse of light activates motor proteins within a region of illumination. Activated motor proteins crosslink the MTs and form a contractile network. Isolation of the network from the solutions requires a second pulse at around $\sim 50-80$s. (b) shows the macroscopic (top row) and microscopic (second row) snapshots of the network, from left to right: during the activation and network formation, at the time of isolation, and shape preserving contraction. The colored dots in the second row track the loci of four distinct microscopic asters in time. The bottom panel shows the profile of the contracting network in time (horizontal axis). The major three phases of the dynamics are separated by vertical lines.
(c) For three different shapes the self-similar contraction of networks is portrayed by overlaying the networks' boundaries as they shrink in time.}
\label{fig1}
\vspace{-3mm}
\end{figure*}

Here, we develop an optical control protocol to activate the motor proteins within a region of illumination, form active MT-motor networks, and isolate them from the surrounding solution. Our strategy utilizes a recently developed optical experimental system to form and isolate active networks of different geometries. Dynamics of the isolated networks are dominated by active stresses with negligible fluid drag; Fig. (\ref{fig1}a) \cite{guntas2015engineering,ross2019controlling}. For a large set of distinct geometries we demonstrate that the active networks undergo shape-preserving contractions. Using a hydrodynamic model we demonstrate that the shape preservation is the direct consequence of viscous-like active stresses. The model teaches us how to program active stresses by modulating the light pattern and intensity. Specifically, we design protocols for spatiotemporal modulations of light intensity to achieve static bending, as well as temporally-alternating bending directions in the network.

\begin{figure*}[ht!]
\centering
\centerline{\includegraphics[width=15cm] {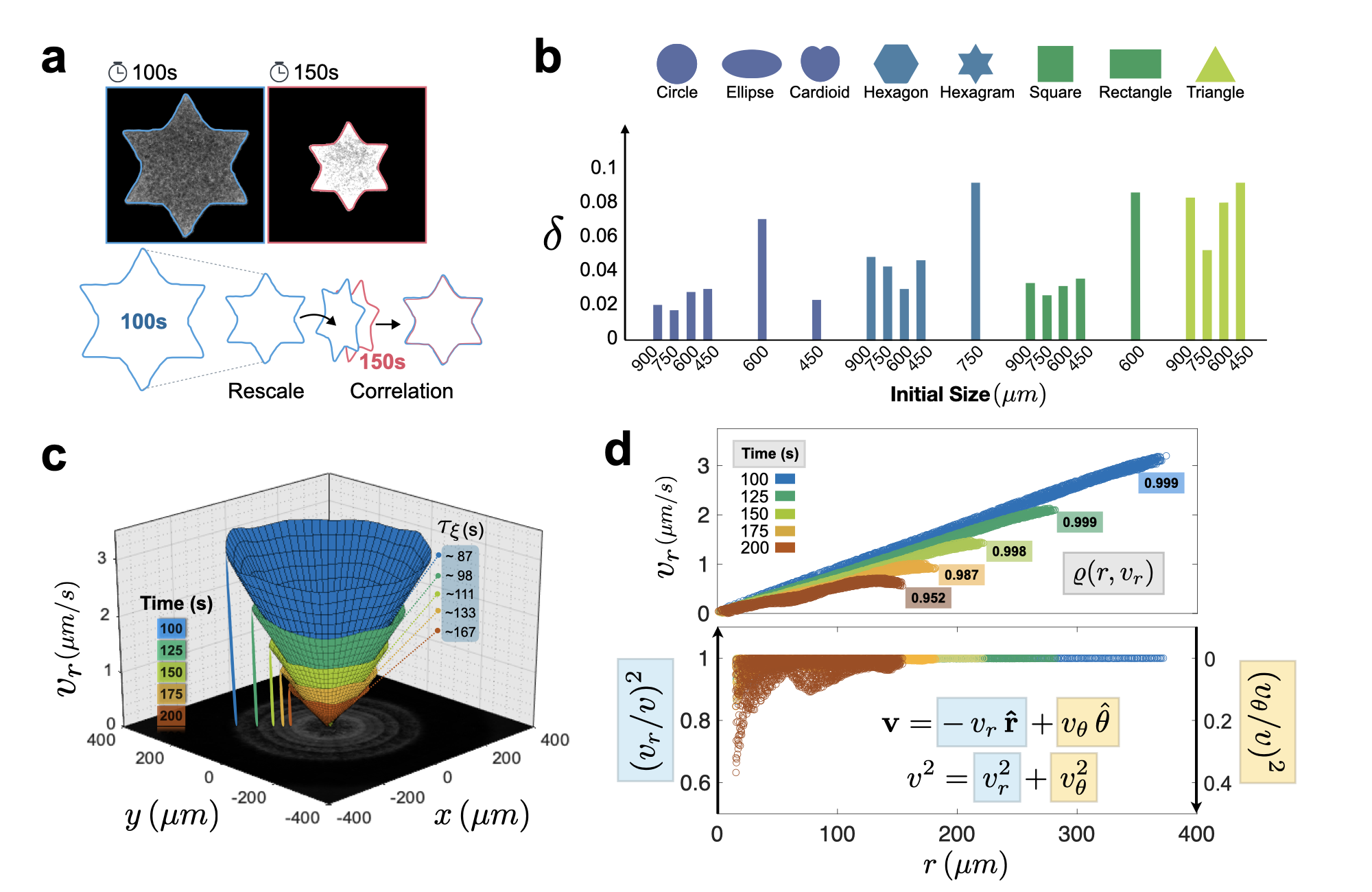}}
\caption{\textbf{Self-similarity persists over time and is the consequence of linear scaling of velocity with radius.} (a) depicts the algorithm for evaluating self-similarity between two timepoints. The boundaries of the two shapes are extracted. The larger shape (earlier time) is scaled down to match the linear scale of the smaller one. The self-similarity is then found by calculating the correlation between the two shapes. (b) For various geometries (and initial sizes) the deviation from self-similarity $(\delta)$ is measured between $t = 100$s and later timepoints $t\leq 360$s (end of contraction). The bars start from zero deviation at $t=100$s and reach their maximum at $t=360$s. All shapes retain their initial shape up to at least $90\%$ accuracy. (c) The magnitude of radial velocity of the contracting networks, at different times. While the radial component increases linearly with distance from the center (hence the cones), the slopes ($\tau_\xi^{-1} = |v_r(r)|/r$) decrease in time. This linearity is assessed in (d): top panel shows the scatter plots of $v_r(r,t)$ vs. $r$. The Pearson correlation coefficient $\varrho(r,v_r)$ is calculated at different timepoints, which varies in time between $0.999$ and $0.952$. The bottom panel shows the relative contribution of the radial and angular components of the velocity. The least contribution of radial component remains above $(v_r/v)^2\simeq 0.65$.
}
\label{fig2}
 \vspace{-3mm}
\end{figure*}

\subsection*{Activity preserves the shape memory of contracting networks}
Performing experiments on several distinct geometries reveals striking universal dynamics that shed light on the underlying active mechanism. We first studied contracting circles as well as polygonal networks (squares, triangles, and hexagons) of different sizes; $460,\,600,\,750$ and $900{\mu}{\text{m}}$. We used a combination of microscopy and image analysis to track and infer network dynamics using labeled MTs.

We found that across a wide range of geometries the MT-motor networks generate a contraction that is self-similar, i.e. shape preserving. We realized that the dynamics of networks consist of three phases: (I) Formation of MT-motor contractile networks, the shapes of which are determined by the region of illumination. The activated network is isolated from the background solution by the end of this phase. (II) Contraction phase during which the area of the network decreases over time while density of cross-linked network increases. (III) Deceleration of contraction as the density of filaments, and thus the MT-MT steric interactions increase. During the contractile phases (II and III), the network retains the initial shape of the light pattern. 

In order to assess self-similarity, we first segment images to find the regions occupied by the networks at different times. Next, for two shapes at timepoints $t_1$ and $t_2>t_1$, with areas $A_1$ and $A_2<A_1$, we scale down the larger shape by $\sqrt{A_2/A_1}$, and align the centers of the two shapes. Self-similarity is defined as the ratio of the bitwise overlap (AND operator) area, and $A_2$ (Fig. (\ref{fig2}a)). To account for stochastic rigid rotations of each network around its center of mass, we maximize the self-similarity with respect to relative rotations over the range of $(-20\,,+20)$ degrees. The deviation from self-similarity, $\delta(t_1,t_2)$, is calculated by subtracting the self-similarity from unity. Across all networks examined we found that $\delta \lesssim 10\%$ over the entire course of the dynamics; Fig. (\ref{fig2}b).

The self-similar scaling of the network boundary over time is strongly suggestive of an underlying contractile mechanism that is distinct from those in passive systems. In a passive system, competition between bulk and boundary energies, along with the dissipative drag forces induced by the fluid, lead to distortions in the curvature of the initial network boundary that increase in time. The absence of these ``equilibrating" (stress releasing) deformations in our system is indicative of strongly activity-driven dynamics, counteracting the dissipative effects.

In comparison to convex shapes that are identified by uniformly positive boundary curvature, the richer geometric features of concave shapes (arcs of positive and negative curvatures) make the deviations from self-similar contraction easier to detect. Furthermore, boundary deformations in concave shapes are more probable to occur due to the bulk-boundary couplings, making the dynamics of concave shapes more informative from a physical perspective. Passive systems with free boundaries equilibrate to round shapes to minimize the sum of the bulk and boundary free energy, and perturbing the boundaries induces stresses in the bulk. Therefore, probing concave active networks provides a more stringent test for verifying the activity-dominated and drag-free contraction.

We prepared networks in two concave geometries: hexagrams and cardioids, and found that these shapes contract with self-similarities indistinguishable from those generated in convex networks. In Figure (\ref{fig2}b), we show for all shapes, the maximum deviation from self-similarity over the course of contraction measured with respect to the reference time $t=100$s. The deviation from self-similar contraction remains below $10\%$ for all convex shapes---in many cases below $5\%$. Between the two concave shapes, the cardioid shows a very small deviation of $2\%$, the hexagram reaches almost $10\%$ deviation, comparable to triangles and rectangles. The absence of such effects in concave shapes of active networks indicates that the contractile motion of our system is stress-free. More precisely, the contraction corresponds to uniform scaling of the intrinsic metric, in accord with the uniform velocity gradient.

\begin{figure*}[ht!]
\centering
\centerline{\includegraphics[width=13cm] {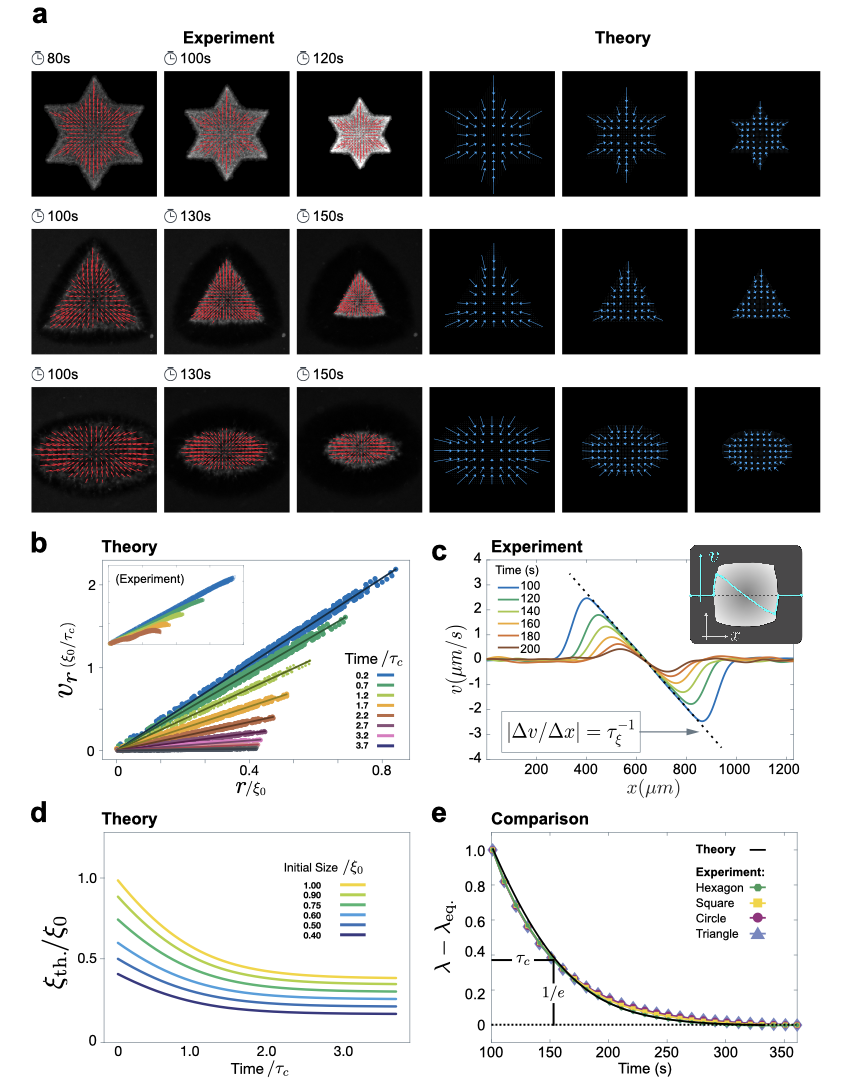}}
\caption{\textbf{Comparison and agreement between experiments and theory supports the role of activity-induced viscous interaction in shape-preserving contraction.} (a) For three shapes of hexagram, triangle and ellipse the velocity vector fields are shown as extracted via PIV in experiments (red) and simulated (blue); note the radial velocity in all cases. (b) The theoretical counterpart of the previously shown $v_r$ vs. $v$ experimental data in Fig. (\ref{fig2}d). (c) For closer comparison the velocity along $x-$axis is plotted at different times for a contracting square. (d) The linear length scale $\xi_{\text{th.}}$ in time as predicted by theory, for different initial sizes. In (e) the fractional contraction $\lambda - \lambda_{\text{eq.}}$ for theoretical results collapse onto a single curve that is compared with those extracted from experiments. The curves decay exponentially over timescale $\tau_c$.
}
\label{fig3}
 \vspace{-3mm}
\end{figure*}

\subsection*{Persistent self-similarity suggests linear radial velocity field}
High degree of persisting shape preservation suggests spatially-uniform and isotropic contraction of the networks. In accord with self-similarity, we found that the contracting networks generate a velocity field, as inferred from Particle Image Velocimetry (PIV), that remains linear throughout the dynamics across all network shapes and sizes. Specifically, Fig. (\ref{fig2}c) shows the radial component of velocity field in the $x-y$ plane, generated by a contracting circle at different times. Similarly in Fig. (\ref{fig2}d) top panel, the radial velocity is plotted as a function of distance $r$ from the center of mass. Linearity of velocity is evident from the Pearson correlation coefficient $\varrho(r , v_r)$ which remains very close to unity. The slope of $v_r$ vs. $r$, corresponding to $\tau_{\xi}^{-1}$ in Fig. (\ref{fig2}c), changes as a function of time and size of the network, hence the subscript $\xi$. The inverse of this slope ($\tau_{\xi}$) can be interpreted as the time it takes for a network of size $\xi$ to shrink to zero, if the contraction would not decelerate. However the contraction of the network leads to accumulation of mass which slows down the contraction, and $\tau_{\xi}$ diverges at an equilibrium density. For a system with free boundary conditions, locally uniform and isotropic contraction implies zero angular velocities. To verify this, we measured the contributions of radial and angular velocity components; Fig. (\ref{fig2}d) bottom panel, and observed that the contribution of angular velocity remains very low for almost the entire course of contraction. In Fig. (\ref{fig2}c), for visual clarity, we only show the velocity cones for a circle. However, the linearity of velocity as a function of distance, and the decrease of the slopes in time, hold true across all networks with different shapes and sizes.

\subsection*{Hydrodynamic model reveals mechanism of universality of self-similar contraction}
Programming active contractile networks requires quantitative understanding of the response of the system to the external probes, e.g. light in our experiments. To understand how self-similar contractions emerge in response to internally generated stress, we developed and analyzed a coarse-grained hydrodynamic model of active networks. Our phenomenology draws on the following experimentally grounded postulates: (1) Isotropicity: the initially randomly oriented MTs organize small asters that are connected to each other via some intermediate MTs. The asters are, however, connected in random directions. Therefore for length scales of multiple asters size isotropicity seems to be a reasonable assumption; see the zoomed panels in Fig. (\ref{fig1}b). (2) Activated motor proteins induce contractile stress. (3) Steric interactions become progressively stronger as the network contracts, and balance out the contractile stress at an equilibrium density of the network.

The hydrodynamics of the system is governed by the conservation laws of total mass and momentum, where total refers to the MT network and the fluid. Mass conservation demands $\partial_t(\rho_{n}+\rho_{f}) = - \nabla\cdot(\rho_{n}\vv_n + \rho_f \vv_f ) = 0$, where $\partial_t$ denotes the partial time derivative, and $\rho_{n/f}$ are network/fluid densities. We drop the network's subscript hereafter. Neglecting the inertial terms on macroscopic time scales, momentum conservation (force balance) for the network requires $\nabla\cdot \bm\sigma^p = \gamma (\vv - \vv_f).$ Here $\nabla\cdot\bm\sigma^p$ is the passive external force exerted from the surrounding fluid on the network, and $\gamma$ is the effective drag coefficient. On the other hand, the viscoelastic response of the network to the total stress reads $\bm\sigma^p + \bm\sigma^a = \eta\nabla\vv$, in which $\bm\sigma^a$ is the active stress, and $\eta$ is the effective network viscosity. Under the assumption of $|\vv_f|\ll|\vv|$, we get
\begin{subequations}{\label{eq:main}}
\begin{align}
&\partial_t\,\rho + \nabla\cdot(\rho\vv) = 0,    \\
&\eta\nabla^2\vv - \gamma\vv = \nabla\cdot\bm\sigma^a.
\end{align}
\end{subequations}

The dependency of the active stress on the intensity of light is crucial to programming the dynamics of network. In order to understand this dependency we simulate the dynamics of contractile networks based on the following assumptions, and assess their validity by comparing the results against experiments. Active stress is assumed to be isotropic, namely proportional to the identity matrix $\mathbb{1}$. In 2D we have $\bm\sigma^a = \frac{1}{2}\text{tr}({\bm\sigma^a})\mathbb{1}\equiv \sigma^a\mathbb{1}$. The active stress can be decomposed into two opposing terms: a contractile term $(\propto \rho)$, and an expansile steric term $(\propto \rho^2)$. Strictly speaking, steric interactions are not intrinsically active, but emerge due to the activity-induced compression. The proportionality constants are assumed to increase linearly with the density of activated motor proteins, in turn an increasing function of the light intensity. The competition between the contractile and the steric interactions vanishes at an equilibrium density $\rho_{\text{eq.}}$, corresponding to the final size of the network when the contraction stops $\xi_{\text{eq.}} = \xi(t\to\infty)$. 

Simulating the network contraction over a range of convex and concave shapes we observe self-similar contractions across all geometries. In the activity dominated regime, the model yields a linear velocity field whose magnitude scales linearly with the distance from the network's center of mass. Specifically, the ratio $\gamma\xi^{2}/\eta$ specifies the relative magnitude of passive and active forces over the longest contractile mode of contraction. In the high-activity regime, the model asymptotically reduces to $\eta \nabla\mathbf{v} = \bm{\sigma}^a$, and the velocity field can be solved given a MT network density. For a network of instantaneous size $\xi(t)$, with uniform MT density and free boundary conditions, the solutions of Eqs. (\ref{eq:main}) are radially symmetric vector fields with constant radial gradient of the form $|\nabla \mathbf{v}(t)| = |v(r=\xi(t))|/\xi(t) = |\sigma^a(t)|/\eta$.

The linearity of the radial vector field $\vv = -\sigma^a\,\rr/\eta$ persists over the course of dynamics, when the two Eqs. (\ref{eq:main}) are solved simultaneously. As such, the velocity field generates angle-preserving dynamics: given points $\xx_1$ and $\xx_2$ in material (Lagrangian) coordinates of the network, their relative position vectors in Eulerian description is scaled by a factor that depends on the time points $t,s$, such that $\left[\rr(\xx_1,t)-\rr(\xx_2,t)\right] \propto \left[\rr(\xx_1,s)-\rr(\xx_2,s)\right]$. Thus, the linear velocity field generated in the activity-dominated regime, induces a self-similar, distance scaling map.

\begin{figure*}[ht!]
\centering
\centerline{\includegraphics[width=15cm] {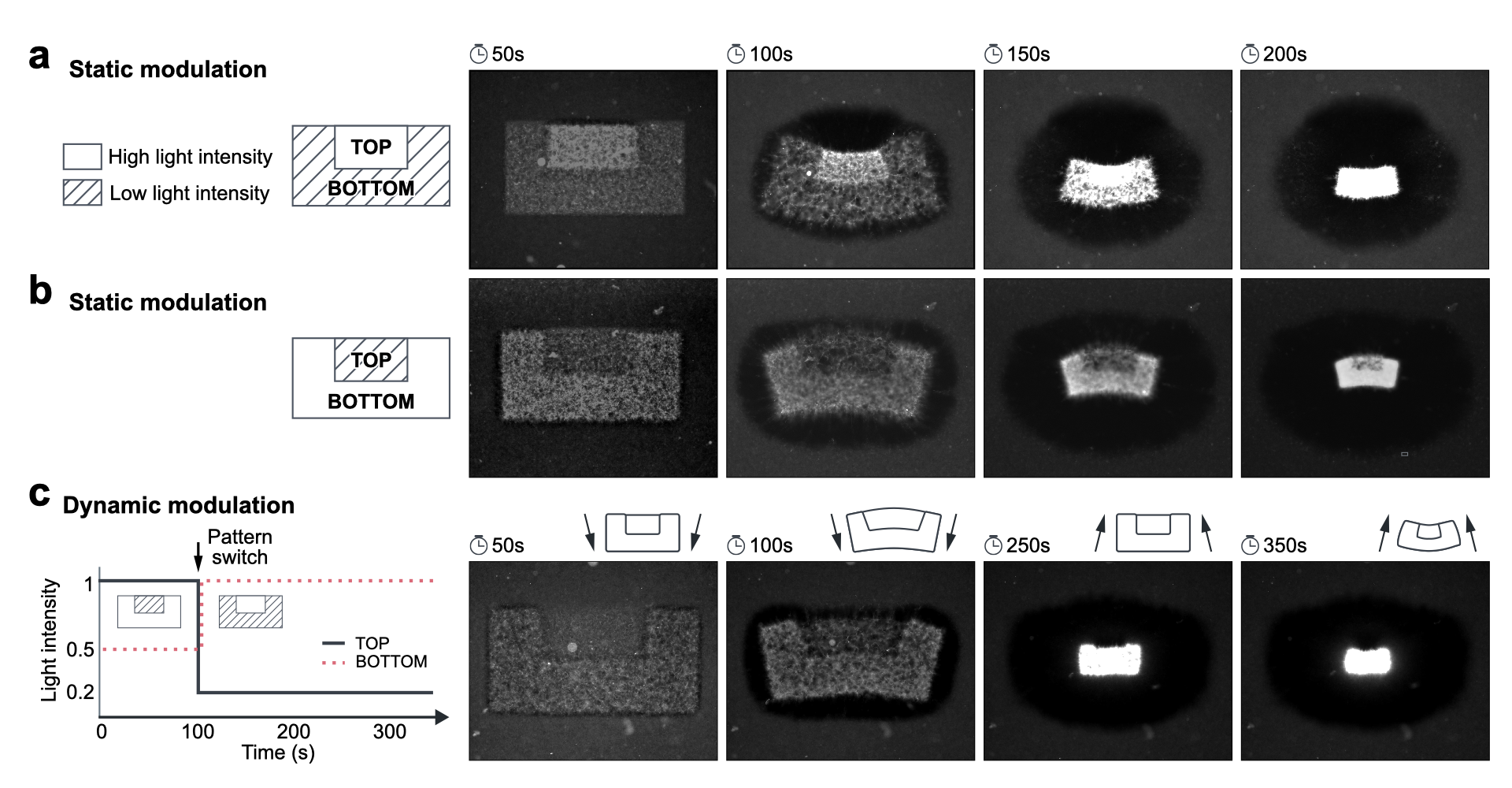}}
\caption{\textbf{Contraction of the network can be programmed via modulating the pattern of illumination in space and time.} (a)/(b) show purely-spatial modulations of light, where the top segment of the rectangle is illuminated more/less strongly. Greater intensity of light activates larger number of motor proteins and thus generates larger active stresses, which leads to larger and faster contraction on the brighter side, and causes bending. In (c) the pattern of illumination varies in time to interpolate between the two static patterns of (a) and (b). Using this dynamic modulation we manage to change the bending direction as the network contracts.}
\label{fig4}
 \vspace{-3mm}
\end{figure*}

Linear velocity field with density-dependent gradient leads to universal dynamics.
The density dependence of the velocity field appears through the stress tensor which determines the instantaneous slope. Active stress $\sigma^a$ is proportional to (a) $\rho - \rho_{\text{eq.}}$, and (b) activity $a_{\ii}$, determined by the concentration of activated motor proteins, assumed to be proportional to the light intensity $\ii$. Together with continuity equation, our model suggests a universality in the velocity field across different shapes and initial conditions. For linear contractions, the density of the MT network remains uniform during the initial phases of the contraction. From continuity equation the density of the network can be expressed as a function of contracted area as $\rho(t)\xi^{2}(t) = \rho_{0} \xi_{0}^2 = \rho_{\text{eq.}} \xi_{\text{eq.}}^2$. Here $t_0$ is a reference time at which the density equals $\rho_0$. Combined with momentum conservation which determines the velocity field, we obtain: $\rho(t) = \rho_{\text{eq.}} + (\rho_0-\rho_{\text{eq.}})\exp(-2(t-t_0)/\tau_c)$, where $\tau_c$ is the contraction timescale and can be expressed in terms of model parameters $\tau_c = \eta/a_{\ii}$; i.e. inversely proportional to activity. Correspondingly the linear size of the network can be expressed as $\xi(t) = \xi_{\text{eq.}} + (\xi_0 - \xi_{\text{eq.}})\exp(-(t-t_0)/\tau_c)$. The normalized fractional contraction thus follows an exponential decay of the form $\lambda(t) - \lambda_{\text{eq.}} = \exp(-(t-t_0)/\tau_c)$.

The results of the simulations for all shapes reproduce the same dynamics as observed in experiments, specifically for $\mathbf{v}(r,t)$ and $\lambda(t)$. The velocity field extracted by PIV from contracting networks, and those obtained from simulations are both linear and radial over shapes and over time; see Fig. (\ref{fig3}a) for qualitative comparison. Consistent with experiments, we observe a linear velocity field over complete contractile dynamics for all shapes analyzed, and the divergence/slope of the velocity field decreases with decreasing size, or equivalently increasing density; Fig. (\ref{fig3}b,c). In our experiments, we held the initial MT density constant across networks of different sizes and shapes. The fractional contraction for experiments on several shapes, as well as those obtained from the theory, as plotted in Fig. (\ref{fig3}d), collapse onto the an exponential curve with decay time of $\tau_c$ which is inversely proportional to the activity. Given that activity is an increasing function of the light intensity, we expect the contraction to speed up upon cranking up the intensity.

\subsection*{Programming deformation through spatial and temporal modulation of activity}
The hydrodynamic model suggests a simple strategy for programming the mechanical properties of MT networks through spatial-temporal modulation of activity. In our hydrodynamic model, the divergence/slope of the contractile velocity field depends on MT density and the activity $a_{\ii}$, which sets the magnitude of stress and thus the contraction timescale. Activity can be modulated experimentally in time and space with light, providing a mechanism to modulate the mechanical behavior of the networks. Spatially-uniform illumination induces uniform activity and isotropic stress which leads to shape-preserving contraction. However, modulation of light pattern in space can generate a nonuniform stress tensor which leads to network regions that contract at different rates. By modulating light levels we modulate the relative local contraction rates which no longer preserve the shape of the network.

Specifically, we generated networks where spatially distinct regions experience two different light intensities and, thus, generate two different contractile fields in close proximity. Differing contractile forces along the boundary between the two networks lead to deformation and bending. Thus, by modulating relative activity, we can induce deformation along the boundary of a network and program novel mechanical behaviors that deviate from the self-similar contractions observed in networks at uniform activity. 

We created a series of light patterns that modulate the relative activity to induce bending deformations. For example, we created a hinge pattern where distinct contractile networks are separated by a joint region, and in the joint region differences in activity lead to relative differences in contractile velocity fields and network bending; Fig. (\ref{fig4}a). In a complement hinge pattern, we induce bending along the opposite direction by switching the orientation of the joint, see Fig. (\ref{fig4}b). 

In addition to generating static deformations, spatial and temporal modulation of light patterns allow the generation of dynamical contraction and deformation through temporal modulation of relative activity. In particular, we temporally modulated the relative light intensity in the two regions of the hinge according to the following protocol. First we shine a light pattern that induces downward bending. The light pattern is subsequently swapped to the complementary pattern at around $t=100$s after the initial illumination. The differential intensities lead to reversal of the bending direction. The rates of the bending and reversal depend on the relative sizes of the two regions of illumination, relative light intensities, and the time at which swapping to complementary pattern takes place. Here we chose a relatively straightforward protocol with the same intensities and densities of MTs as chosen in the previously discussed case of self-similar contractions. 

Broadly, these experiments show that both spatial-temporal modulation of light intensity allows us to induce programmed patterns of mechanical deformation into active MT networks. In this way, the natural shape preservation property of active MT networks can be simply modulated through relative differences in activity in distinct parts of an induced network. This controllability of MT networks allows us to program units of networks in which different possess engineered mechanical properties and can perform work in a programmed and predetermined manner through internal couplings. 

\subsection*{Discussion}
Active networks are ubiquitous in biology, and their non-equilibrium properties are poorly understood. Our work reveals signature of activity in the mechanical properties at macroscopic scales. The self-similar contraction is intrinsically related to the non-equilibrium nature of the system, which preserves a geometric memory, unlike in passive systems where equilibration increases entropy and erases the memory of the initial state. This memory preservation property makes the behavior of the system more controllable without the need to tuning the microscopic degrees of freedom. 

Previous works analyzed active contractions in networks of MT and actin in cell extracts, where the contracting network is embedded in a viscous solution, thus subjected to drag forces. Our optical control strategies allow us to isolate the networks from passive boundaries while using light to modulate the shape and activity. Further, in conventional materials altering mechanical properties requires changing the microscopic structure of the material, for example, through doping. These changes are generically irreversible (plastic), and are hard to be modulated at the microscopic level. In our systems, the degree of linking of the network and the active stresses can be tuned in space and time, enabling a separate strategy for the programming and control over material mechanics. Activity induced deformations provide a strategy for engineering novel behaviors at micron length scales.

\begin{acknowledgments}
The authors are grateful to Inna-Marie Strazhnik for making illustrations, and to John Brady, Dominik Schildknecht and Enrique Amaya Perez for useful discussions. MT was supported by Packard Foundation, Rosen Center for Bioengineering, and Heritage Medical Research Institute. RP was supported by NIH grant number 1R35 GM118043-01. RP and MT would like to thank Foundational Questions Institute and Fetzer Franklin Fund through FQXi 1816 for funding the research.
\end{acknowledgments}

\appendix
\section{Instrumentation and Imaging}{\label{sec:methods}}
\subsection{Active Matter System and Sample Chambers}
The system consists of stabilized microtubule, kinesin motors (constructed with light-induced hetero-dimer system) and an energy mix. All ingredients and buffer preparation protocol are documented in a previous paper by Ross et. al.\cite{ross2019controlling}, and we follow the exact same procedure in our study. The sample chambers are made by sandwiching pre-cut Parafilm M by coated slides and coverslips \cite{ross2019controlling,Lau_acrylamidecoat2009}. The measured depth of the chamber is approximately 80$\mu$m.

\subsection{Microscope Instrumentation}
The experiment is conducted on an epifluorescence microscope (Nikon Ti2) with 10X magnification (Nikon Plan Apo $\lambda$ 10X). We customize the system by adding a programmable digital light projector (EKB Technologies DLP LightCrafter E4500 MKII Fiber Couple), which is used to image the light pattern activating the dimerization of kinesin motors. The DLP chip is illuminated by the four-channeled LED (ThorLabs LED4D067) at the wavelength of 470nm. Fluorescently labeled microtubules are illuminated by 660nm and imaged with digital camera (Hamamatsu orca-flash 4.0). The system is controlled with Micro-Manager on PC.

\subsection{Control Strategy for Isolating the Contracting Network}
When the light patterns are constantly projected onto the reaction sample, the contraction is accompanied by the formation of canals at the sharp corners of the pattern (e.g. vertices of polygons). These canals pave paths for the background solution---containing floating Mts---to pour into the region of illumination. These MTs get cross-linked upon entering this region, and form a steady state of flow; hence coupling of network and the background fluid. To isolate the cross-linked network from the ambient solution, we decrease the size of the projected pattern to prevent new MT-solution mix from flowing in. As shown in Fig.(\ref{fig1}a), we first projected a pattern at full size to initiate the network cross-linking. After 80s, a shrunken pattern is projected, with the same geometry and light intensity, but with 70\% linear size of the initial pattern. After this initial phase, the sample is constantly illuminated every 10s with 30ms duration, during which the light pattern is further decreased to 50\% original linear size gradually over the course of contraction which stops at 5min.

\section{Image Processing}

\subsection{Segmentation and Detection of the Network}
The time lapse images of contracting network is segmented and isolated from the background solution utilizing a few built-in function of MATLAB Image Processing toolbox. During the contraction (phase II) the boundaries of the network is well separated from the solution that allows for segmentation. The steps are as follows: We first subtract the local background intensity using {\textsf{imflatfield}} function over regions of sizes of $\sim 300\,({\mu}m)$. This is required to remove artificial shadows. Next we use the watershed algorithm to separate the network from the background fluid.\\ 
\subsection{Measuring Velocity and Density Fields}
Velocity field is extracted at different time points using the built-in MATLAB function {\textsf{imregtform}}. This function estimates the displacement field $\mathbf D_{1\to 2}$ that warps the images at times $t_1$ onto the image at $t_2$. In the Lagrangian picture for a point labeled by $\mathbf p$, we get $\rr(\mathbf p, t_2) = \rr(\mathbf p, t_1) + \mathbf D_{1\to 2}(\mathbf p)$. The displacement field is then converted to our units using the pixel value of $0.65 \mu$m. We define the velocity field in terms of $\overline{\rr}$ and $\overline t$, where 
\begin{align}
    \overline{\rr} = \frac{1}{2}\,\bigg(\rr(\mathbf p , t_2) + \rr(\mathbf p , t_1)\bigg),\qquad
    \overline{t} = \frac{1}{2}\,\big(t_1 + t_2\big).
\end{align}
The velocity field reads
\begin{equation}
    \vv(\overline{\rr} , \overline{t}) = \mathbf D_{1\to 2}(\mathbf p) \times 0.65/(t_2 - t_1)
\end{equation}

In order to measure the velocity and density as a function of distance from center of mass (CoM), the center of mass of the network is found at each time point. Under the assumption that the local density of the network $\rho(\rr)$, is proportional to the intensity of light captured in gray-scale images $\ii(\rr)$, the center of mass is obtained by 
\begin{equation}{\label{eq:CoM}}
    \RR(t) = \frac{\int_{\text{net.}}d^2r\; \rr\, \ii(\rr) }{\int_{\text{net.}}d^2r\; \ii(\rr)},
\end{equation}
where $\int_{\text{net.}} d^2r$ integrates over the area of the network. 

For later time points when the intensity is saturated; hence not proportional to density, an alternative method is to use the velocity field of the network to estimate the position (and velocity) of the CoM. From Eq. (\ref{eq:CoM}), we have:
\begin{equation}
    \mathbf{V}_{\text{CoM}}(t) = M^{-1}\,\int_{\text{net.}}d^2r\; \vv\, \rho(\rr).
\end{equation}
Here $M$ is the total mass of the network---assumed to be conserved during the course of contraction; thus calculable from earlier time points when the density is safely assumed to be proportional to intensity. Redefining the position vector and velocities relative to those of the CoM we get $\rr \equiv \rr - \RR$; and $\vv(\rr , t) \equiv \vv(\rr,t) - \vv(\RR , t)$. Note that although on average the CoM is stationary, on the short timescales it is subject to small and fast random fluctuations due to the noisy background flows. The redefinition of velocity field ensures $\int_{\text{net.}}d^2r\,\vv(\rr)\,\ii(\rr) = 0$. Therefore the CoM is now determined as the point at which the relative velocity vanishes. To find the velocity as a function of $|\rr|$ (from CoM), the magnitude of the relative velocity is averaged over all points at a radius.

\bibliographystyle{unsrt}
\bibliography{active_contraction.bbl}

\begin{thebibliography}{10}

\bibitem{marchetti2013hydrodynamics}
M~Cristina Marchetti, Jean-Fran{\c{c}}ois Joanny, Sriram Ramaswamy,
  Tanniemola~B Liverpool, Jacques Prost, Madan Rao, and R~Aditi Simha.
\newblock Hydrodynamics of soft active matter.
\newblock {\em Reviews of Modern Physics}, 85(3):1143, 2013.

\bibitem{ramaswamy2010mechanics}
Sriram Ramaswamy.
\newblock The mechanics and statistics of active matter.
\newblock {\em Annual Review of Condensed Matter Physics}, 2010.

\bibitem{bechinger2016active}
Clemens Bechinger, Roberto Di~Leonardo, Hartmut L{\"o}wen, Charles Reichhardt,
  Giorgio Volpe, and Giovanni Volpe.
\newblock Active particles in complex and crowded environments.
\newblock {\em Reviews of Modern Physics}, 88(4):045006, 2016.

\bibitem{burla2019mechanical}
Federica Burla, Yuval Mulla, Bart~E Vos, Anders Aufderhorst-Roberts, and
  Gijsje~H Koenderink.
\newblock From mechanical resilience to active material properties in
  biopolymer networks.
\newblock {\em Nature Reviews Physics}, 1(4):249--263, 2019.

\bibitem{peraza2014origami}
Edwin~A Peraza-Hernandez, Darren~J Hartl, Richard~J Malak~Jr, and Dimitris~C
  Lagoudas.
\newblock Origami-inspired active structures: a synthesis and review.
\newblock {\em Smart Materials and Structures}, 23(9):094001, 2014.

\bibitem{pinson2017self}
Matthew~B Pinson, Menachem Stern, Alexandra~Carruthers Ferrero, Thomas~A
  Witten, Elizabeth Chen, and Arvind Murugan.
\newblock Self-folding origami at any energy scale.
\newblock {\em Nature communications}, 8(1):1--8, 2017.

\bibitem{furthauer2020design}
Sebastian F{\"u}rthauer, Daniel~J Needleman, and Michael~J Shelley.
\newblock A design framework for actively crosslinked filament networks.
\newblock {\em arXiv preprint arXiv:2009.09006}, 2020.

\bibitem{nedelec1997self}
FJ~Nedelec, Thomas Surrey, Anthony~C Maggs, and Stanislas Leibler.
\newblock Self-organization of microtubules and motors.
\newblock {\em Nature}, 389(6648):305, 1997.

\bibitem{surrey2001physical}
Thomas Surrey, Fran{\c{c}}ois N{\'e}d{\'e}lec, Stanislas Leibler, and Eric
  Karsenti.
\newblock Physical properties determining self-organization of motors and
  microtubules.
\newblock {\em Science}, 292(5519):1167--1171, 2001.

\bibitem{sanchez2012spontaneous}
Tim Sanchez, Daniel~TN Chen, Stephen~J DeCamp, Michael Heymann, and Zvonimir
  Dogic.
\newblock Spontaneous motion in hierarchically assembled active matter.
\newblock {\em Nature}, 491(7424):431--434, 2012.

\bibitem{foster2015active}
Peter~J Foster, Sebastian F{\"u}rthauer, Michael~J Shelley, and Daniel~J
  Needleman.
\newblock Active contraction of microtubule networks.
\newblock {\em Elife}, 4:e10837, 2015.

\bibitem{mizuno2007nonequilibrium}
Daisuke Mizuno, Catherine Tardin, Christoph~F Schmidt, and Frederik~C
  MacKintosh.
\newblock Nonequilibrium mechanics of active cytoskeletal networks.
\newblock {\em Science}, 315(5810):370--373, 2007.

\bibitem{thoresen2011reconstitution}
Todd Thoresen, Martin Lenz, and Margaret~L Gardel.
\newblock Reconstitution of contractile actomyosin bundles.
\newblock {\em Biophysical journal}, 100(11):2698--2705, 2011.

\bibitem{kohler2011structure}
Simone K{\"o}hler, Volker Schaller, and Andreas~R Bausch.
\newblock Structure formation in active networks.
\newblock {\em Nature materials}, 10(6):462--468, 2011.

\bibitem{kohler2012contraction}
Simone K{\"o}hler and Andreas~R Bausch.
\newblock Contraction mechanisms in composite active actin networks.
\newblock {\em PloS one}, 7(7):e39869, 2012.

\bibitem{schuppler2016boundaries}
Matthias Schuppler, Felix~C Keber, Martin Kr{\"o}ger, and Andreas~R Bausch.
\newblock Boundaries steer the contraction of active gels.
\newblock {\em Nature communications}, 7(1):1--10, 2016.

\bibitem{foster2017connecting}
Peter~J Foster, Wen Yan, Sebastian F{\"u}rthauer, Michael~J Shelley, and
  Daniel~J Needleman.
\newblock Connecting macroscopic dynamics with microscopic properties in active
  microtubule network contraction.
\newblock {\em New Journal of Physics}, 19(12):125011, 2017.

\bibitem{suzuki2017spatial}
Kazuya Suzuki, Makito Miyazaki, Jun Takagi, Takeshi Itabashi, and Shin’ichi
  Ishiwata.
\newblock Spatial confinement of active microtubule networks induces
  large-scale rotational cytoplasmic flow.
\newblock {\em Proceedings of the National Academy of Sciences},
  114(11):2922--2927, 2017.

\bibitem{lee2001macroscopic}
Ha~Youn Lee and Mehran Kardar.
\newblock Macroscopic equations for pattern formation in mixtures of
  microtubules and molecular motors.
\newblock {\em Physical Review E}, 64(5):056113, 2001.

\bibitem{liverpool2003instabilities}
Tanniemola~B Liverpool and M~Cristina Marchetti.
\newblock Instabilities of isotropic solutions of active polar filaments.
\newblock {\em Physical review letters}, 90(13):138102, 2003.

\bibitem{kruse2004asters}
Karsten Kruse, Jean-Fran{\c{c}}ois Joanny, Frank J{\"u}licher, Jacques Prost,
  and Ken Sekimoto.
\newblock Asters, vortices, and rotating spirals in active gels of polar
  filaments.
\newblock {\em Physical review letters}, 92(7):078101, 2004.

\bibitem{aranson2005pattern}
Igor~S Aranson and Lev~S Tsimring.
\newblock Pattern formation of microtubules and motors: Inelastic interaction
  of polar rods.
\newblock {\em Physical Review E}, 71(5):050901, 2005.

\bibitem{liverpool2006rheology}
Tanniemola~B Liverpool and M~Cristina Marchetti.
\newblock Rheology of active filament solutions.
\newblock {\em Physical review letters}, 97(26):268101, 2006.

\bibitem{juelicher2007active}
Frank Juelicher, Karsten Kruse, Jacques Prost, and J-F Joanny.
\newblock Active behavior of the cytoskeleton.
\newblock {\em Physics reports}, 449(1-3):3--28, 2007.

\bibitem{mackintosh2008nonequilibrium}
Fred~C MacKintosh and Alex~J Levine.
\newblock Nonequilibrium mechanics and dynamics of motor-activated gels.
\newblock {\em Physical review letters}, 100(1):018104, 2008.

\bibitem{koenderink2009active}
Gijsje~H Koenderink, Zvonimir Dogic, Fumihiko Nakamura, Poul~M Bendix,
  Frederick~C MacKintosh, John~H Hartwig, Thomas~P Stossel, and David~A Weitz.
\newblock An active biopolymer network controlled by molecular motors.
\newblock {\em Proceedings of the National Academy of Sciences},
  106(36):15192--15197, 2009.

\bibitem{gao2015multiscale_prl}
Tong Gao, Robert Blackwell, Matthew~A Glaser, Meredith~D Betterton, and
  Michael~J Shelley.
\newblock Multiscale polar theory of microtubule and motor-protein assemblies.
\newblock {\em Physical review letters}, 114(4):048101, 2015.

\bibitem{ronceray2016fiber}
Pierre Ronceray, Chase~P Broedersz, and Martin Lenz.
\newblock Fiber networks amplify active stress.
\newblock {\em Proceedings of the national academy of sciences},
  113(11):2827--2832, 2016.

\bibitem{gladrow2016broken}
J~Gladrow, N~Fakhri, FC~MacKintosh, CF~Schmidt, and CP~Broedersz.
\newblock Broken detailed balance of filament dynamics in active networks.
\newblock {\em Physical review letters}, 116(24):248301, 2016.

\bibitem{furthauer2019self}
Sebastian F{\"u}rthauer, Bezia Lemma, Peter~J Foster, Stephanie~C Ems-McClung,
  Che-Hang Yu, Claire~E Walczak, Zvonimir Dogic, Daniel~J Needleman, and
  Michael~J Shelley.
\newblock Self-straining of actively crosslinked microtubule networks.
\newblock {\em Nature Physics}, 15(12):1295--1300, 2019.

\bibitem{guntas2015engineering}
Gurkan Guntas, Ryan~A Hallett, Seth~P Zimmerman, Tishan Williams, Hayretin
  Yumerefendi, James~E Bear, and Brian Kuhlman.
\newblock Engineering an improved light-induced dimer (ilid) for controlling
  the localization and activity of signaling proteins.
\newblock {\em Proceedings of the National Academy of Sciences},
  112(1):112--117, 2015.

\bibitem{ross2019controlling}
Tyler~D Ross, Heun~Jin Lee, Zijie Qu, Rachel~A Banks, Rob Phillips, and Matt
  Thomson.
\newblock Controlling organization and forces in active matter through
  optically defined boundaries.
\newblock {\em Nature}, 572(7768):224--229, 2019.

\bibitem{Lau_acrylamidecoat2009}
A.~W.~C. Lau, A.~Prasad, and Z.~Dogic.
\newblock Condensation of isolated semi-flexible filaments driven by depletion
  interactions.
\newblock {\em EPL (Europhysics Letters)}, 87(4):48006, 2009.

\end{thebibliography}

\end{document}